\documentclass[twocolumn,floatfix,superscriptaddress,aps,nofootinbib]{revtex4-2}
\usepackage{graphicx}
\usepackage{amsmath}
\usepackage{amssymb}
\usepackage{bm}
\usepackage{hyperref}

\begin{document}

\title{Supercell symmetry modified spectral statistics of Kramers-Weyl fermions}
\author{G. Lemut}
\affiliation{Instituut-Lorentz, Universiteit Leiden, P.O. Box 9506, 2300 RA Leiden, The Netherlands}
\author{M. J. Pacholski}
\affiliation{Instituut-Lorentz, Universiteit Leiden, P.O. Box 9506, 2300 RA Leiden, The Netherlands}
\author{J. Tworzyd{\l}o}
\affiliation{Faculty of Physics, University of Warsaw, ul.\ Pasteura 5,\\
02--093 Warszawa, Poland}
\author{C. W. J. Beenakker}
\affiliation{Instituut-Lorentz, Universiteit Leiden, P.O. Box 9506, 2300 RA Leiden, The Netherlands}
\date{December 2021}
\begin{abstract}
We calculate the spectral statistics of the Kramers-Weyl Hamiltonian $H=v\sum_{\alpha} \sigma_\alpha\sin p_\alpha+t \sigma_0\sum_\alpha\cos p_\alpha$ in a chaotic quantum dot. The Hamiltonian has symplectic time-reversal symmetry ($H$ is invariant when spin $\sigma_\alpha$ and momentum $p_\alpha$ both change sign), and yet for small $t$ the level spacing distribution $P(s)\propto s^\beta$ follows the $\beta=1$ orthogonal ensemble instead of the $\beta=4$ symplectic ensemble. We identify a supercell symmetry of $H$ that explains this finding. The supercell symmetry is broken by the spin-independent hopping energy $\propto t\cos p$, which induces a transition from $\beta=1$ to $\beta=4$ statistics that shows up in the conductance as a transition from weak localization to weak antilocalization.\smallskip\\
\textit{Contribution to the special issue of J.Phys.A in honour of the life and work of Fritz Haake.}
\end{abstract}
\maketitle

\section{Introduction}
\label{intro}

The Wigner surmise $P(s)\propto s^\beta$ for the probability distribution of level spacings \cite{Wig67} is a quantum signature of chaos \cite{Haake}. The exponent $\beta$, the Dyson index \cite{Dys62}, can take on the values 1, 2 or 4, depending on the presence or absence of time-reversal symmetry and spin-rotation symmetry. Electrons in zero magnetic field have $\beta=1$ in the absence of spin-orbit coupling and $\beta=4$ with spin-orbit coupling, while $\beta=2$ in a magnetic field irrespective of the spin degree of freedom. In the context of random-matrix theory one says that the Hamiltonian belongs to the universality class of the Gaussian Orthogonal  Ensemble ($\beta=1$, GOE), Gaussian Unitary Ensemble ($\beta=2$, GUE), or Gaussian Symplectic Ensemble ($\beta=4$, GSE).\footnote{The orthogonal, unitary, and symplectic matrices in this nomenclature refer to the matrix that diagonalizes the Hamiltonian.}

This classification applies both to massive electrons \cite{Guh98} (e.g.\ in a metal grain or in a semiconductor quantum dot) and to massless electrons \cite{Lai18} (e.g.\ in graphene or on the surface of a topological insulator). Here we consider a specific model in the latter category: Massless electrons (Weyl fermions) with a band crossing (Weyl point) enforced by Kramers degeneracy \cite{Cha18,She18}. These low-energy excitations known as Kramers-Weyl fermions appear at time-reversally invariant momenta $\bm{\Pi}$ in the Brillouin zone (such that $\bm{\Pi}$ and $-\bm{\Pi}$ differ by a reciprocal lattice vector). A strong spin-orbit coupling without reflection or mirror symmetry produces a linear band splitting $\pm(\bm{p}-\bm{\Pi})\cdot\bm{\sigma}$ near each of the high-symmetry points. The $\pm$ sign designates the chirality of the excitations. 

On a three-dimensional (3D) cubic lattice (unit lattice constant $a_0$) the Hamiltonian
\begin{align}
H={}& v(\sigma_x\sin p_x+\sigma_y\sin p_y+\sigma_z\sin p_z)\nonumber\\
&+t\sigma_0 (\cos p_x+\cos p_y+\cos p_z)+V(\bm{r})\sigma_0 \label{Hdef}
\end{align}
describes Kramers-Weyl fermions of positive chirality with momenta near $(0, 0, 0)$, $(\pi, \pi, 0)$, $(\pi, 0, \pi)$, $(0, \pi, \pi)$ and of negative chirality near $(\pi, \pi, \pi)$, $(\pi, 0, 0)$, $(0, \pi, 0)$, $(0, 0, \pi)$. The Hamiltonian contains spin-independent terms, hopping terms $\propto \cos p_\alpha$ and a scalar potential $V$, as well as spin-orbit coupling terms $\propto \sigma_\alpha\sin p_\alpha$.

\begin{figure}[tb]
\centerline{\includegraphics[width=0.9\linewidth]{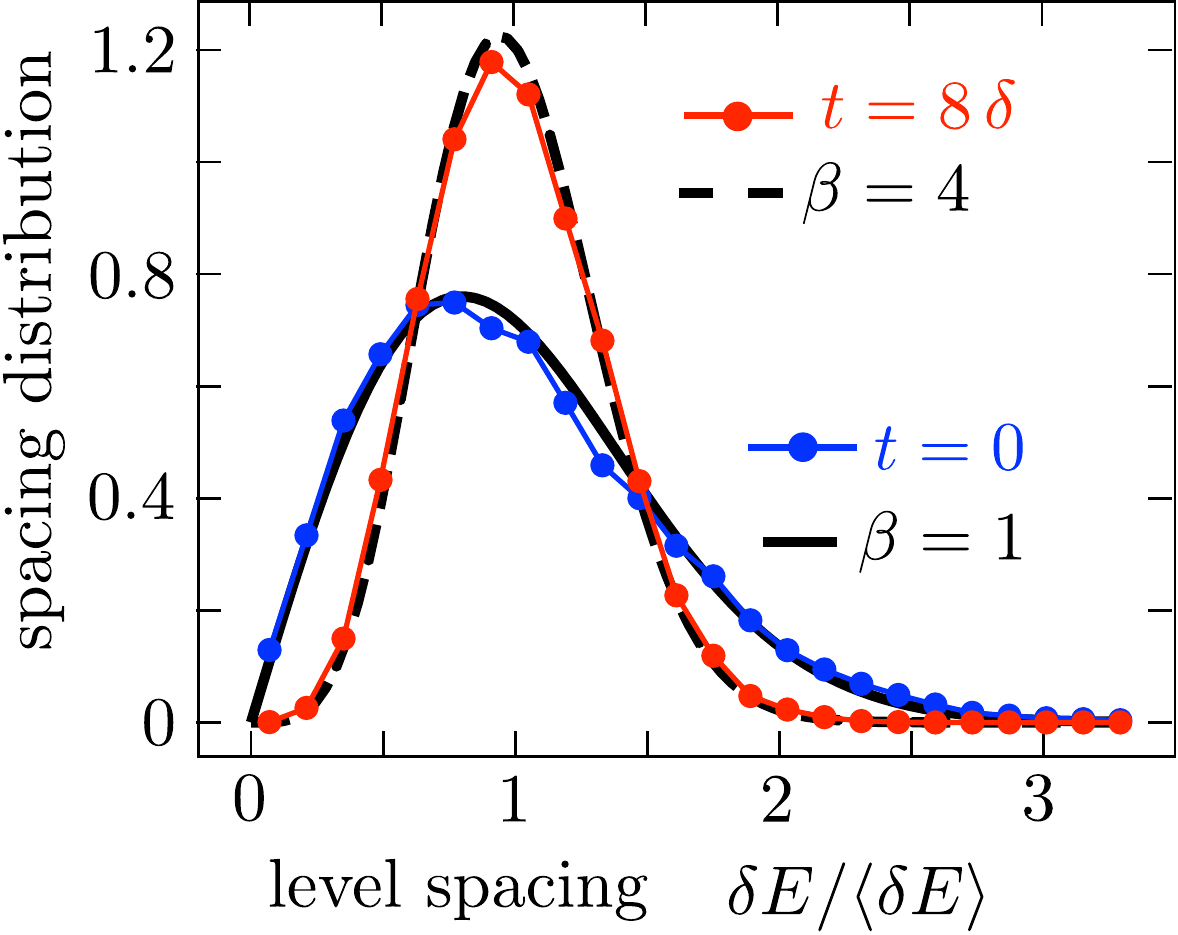}}
\caption{Distribution of the level spacings $\delta E$ (normalized by the mean spacing $\langle\delta E\rangle\equiv\delta=4.04\cdot 10^{-3}\,v/a_0$) of the Hamiltonian \eqref{Hdef}, for $t=0$ (blue) and $t\neq 0$ (red), on a $20\times 20\times 20$ cubic lattice. The potential $V$ was chosen independently on each site from a uniform distribution in the interval $(-V_0/2, V_0/2)$ with $V_0=1.5\,v/a_0$. The solid and dashed black curves give the Wigner surmise for $\beta=1$ and $\beta=4$, respectively.
}
\label{fig_GOE}
\end{figure}

The numerical study of the spectral statistics of Kramers-Weyl fermions that prompted our investigation is shown in Fig.\ \ref{fig_GOE}. A quantum dot is formed by restricting the lattice to a small region and chaotic dynamics is produced by a random potential. For $t=0$ the level spacing distribution is well described by the $\beta=1$ Wigner surmise (orthogonal statistics), while the spin-orbit coupling would have suggested symplectic $\beta=4$ statistics. Paradoxically, the $\beta=4$ distribution requires the addition of spin-\textit{independent} hopping.

In the next section we construct the ``fake'' time-reversal operation ${\cal T}^\ast$ that squares to $+1$ and is responsible for the $\beta=1$ spacing distribution when $t=0$. The supercell symmetry that enables ${\cal T}^\ast$ is broken by the $\cos p$ terms, which reveal the true ${\cal T}$, squaring to $-1$ with a $\beta=4$ spacing distribution. In Sec.\ \ref{WLtoWAL} we investigate how the symmetry breaking manifests itself in a transport property (the magnetoconductance). The analytical results are compared with numerical simulations in Sec.\ \ref{sec_numerics}. In the concluding section we make contact with the spectrum of lattice Dirac operators on a torus, which shows a similar shift of symmetries when the number of lattice sites changes from even to odd \cite{Kie14,Kie17}.

\section{Supercell symmetry}
\label{fake}

\subsection{Zero magnetic field}

The tight-binding Hamiltonian of a spin-1/2 degree of freedom with nearest-neighbor hopping and on-site disorder on an orthorhombic lattice (lattice constants $a_x,a_y,a_z$) has the generic form \cite{Cha18}
\begin{equation}
{\cal H}=\sum_{\alpha=x,y,z}\bigl[t_\alpha\sigma_0\cos a_\alpha p_\alpha+v_\alpha\sigma_\alpha\sin a_\alpha p_\alpha\bigr]+V(\bm{r})\sigma_0.\label{calHdef}
\end{equation}
Both the spin-independent hopping energies $t_\alpha$ and the spin-orbit coupling amplitudes $v_\alpha$ may be anisotropic. We set $\hbar$ equal to unity, $p_\alpha=-i\partial/\partial x_\alpha$ is the momentum operator, the Pauli spin matrices are $\bm{\sigma}=(\sigma_x,\sigma_y,\sigma_z)$, and $\sigma_0$ is the $2\times 2$ unit matrix.

The Hamiltonian \eqref{calHdef} is constrained by the symplectic symmetry
\begin{equation}
{\cal H}=\sigma_y{\cal H}^\ast\sigma_y\equiv {\cal T}{\cal H}{\cal T}.\label{symplectic}
\end{equation}
This is a time-reversal operation that changes the sign of both $\bm{\sigma}$ and $\bm{p}=-i\bm{\nabla}$, leaving ${\cal H}$ invariant. The operator ${\cal T}=\sigma_y\times\text{complex conjugation}$ squares to $-1$, thus we expect GSE statistics, while GOE statistics would require a time-reversal operator that squares to $+1$.

The eight flavors of Kramers-Weyl fermions at $p_\alpha\in\{0,\pi/a_\alpha\}$ are displaced in energy from $E=0$ by the $t_\alpha$ terms. Without these terms, the Hamiltonian
\begin{equation}
{\cal H}_0=\sum_{\alpha}v_\alpha\sigma_\alpha\sin a_\alpha p_\alpha+V(\bm{r})\sigma_0\label{H0def}
\end{equation}
has the supercell symmetry\footnote{The unitary transformation \eqref{supercellsymm} has a periodicity of twice the lattice constant, hence the name ``supercell symmetry'', suggested to us by Anton Akhmerov.}
\begin{equation}
U_y{\cal H}_0 U_y^\dagger={\cal H}_0,\;\;U_y=\sigma_y e^{i\pi n_x+i\pi n_z},\;\;n_\alpha=\frac{x_\alpha}{a_\alpha}\in\mathbb{Z},\label{supercellsymm}
\end{equation}
which transforms $p_x\rightarrow p_x+\pi/a_x$, $p_z\rightarrow p_z+\pi/a_z$, while leaving $p_y$ unaffected. The operator $U_y$ thus maps each Kramers-Weyl fermion onto a partner of the same chirality.

Since $U_y^2=1$ its eigenvalues are $\pm 1$ and we can block-diagonalize ${\cal H}_0$ in sectors of the Hilbert space where $U_y\Psi= \pm \Psi$. In a given sector the time-reversal operator ${\cal T}=\sigma_y\times \text{complex conjugation}$ can be replaced by 
\begin{equation}
{\cal T}^\ast=\pm {\cal T}U_y=\mp e^{i\pi n_x+i\pi n_z}\times\text{complex conjugation}.\label{Tastdef}
\end{equation}
The ``fake'' time-reversal operator ${\cal T}^\ast$ squares to $+1$, so each sector has an orthogonal time-reversal symmetry.

The spin-independent hopping terms in the full Hamiltonian \eqref{calHdef} break the supercell symmetry if two or more of the $t_\alpha$'s are nonzero. (If only a single $t_\alpha\neq 0$ the symmetry $U_\alpha=\sigma_\alpha e^{i\pi\sum_{\alpha'\neq\alpha}n_{\alpha'}}$ remains unbroken.) We would thus expect a $\beta=1$ to $\beta=4$ transition in the level spacing distribution $P(s)\propto s^\beta$ when $t$ becomes larger than the mean level spacing $\delta$.

\subsection{Nonzero magnetic field}

A magnetic field $B$ breaks time-reversal symmetry, driving both orthogonal ($\beta=1$) and symplectic ($\beta=4$) level spacing distributions towards the unitary ($\beta=2$) result. The degeneracy of the $\beta=2$ spectra is different in the two cases.

For $B=0$ each energy level is twofold degenerate (Kramers degeneracy). In a magnetic field the degeneracy is broken for a nonzero $t_\alpha$, but it remains when $t_x,t_y,t_z=0$ if the magnetic field enters only via the substitution $\bm{p}\rightarrow \bm{p}+e\bm{A}$ --- so only as an orbital effect, no Zeeman effect on the spin.

This persistent degeneracy is due to the fact that the supercell symmetry $U_\alpha$ is not broken by the substitution $\bm{p}\rightarrow \bm{p}+e\bm{A}$. Starting from a Hamiltonian which commutes with $U_x$ and $U_y$ and an energy eigenstate $\Psi$ such that $U_y\Psi=\Psi$ we can then construct another eigenstate $\Psi'=U_{x}\Psi$ at the same energy eigenvalue. The two states $\Psi$ and $\Psi'$ are orthogonal,
\begin{align}
\langle\Psi|\Psi'\rangle&=\langle\Psi|U_x|\Psi\rangle=\langle\Psi|U_y^\dagger U_xU_y|\Psi\rangle\nonumber\\
&=-\langle\Psi|U_x|\Psi\rangle=-\langle\Psi|\Psi'\rangle\Rightarrow\langle\Psi|\Psi'\rangle=0,
\end{align}
so the energy eigenvalue is twofold degenerate.

\section{Supercell symmetry effects on the conductance}
\label{WLtoWAL}

The appearance of the supercell symmetry can be probed via the electrical conductance $G$. In a magnetic field, the $\beta=1\rightarrow\beta=2$ transition gives an increase in $G$ (weak localization), while the $\beta=4\rightarrow\beta=2$ transition gives a decrease in $G$ (weak antilocalization). The theoretical prediction for this quantum correction $\delta G=G(B)-G(0)$ is \cite{Bee97}
\begin{equation}
\delta G=\frac{2e^2}{h}\times\begin{cases}
1/3&\text{for}\;\;\beta=1\rightarrow 2,\\
-1/6&\text{for}\;\;\beta=4\rightarrow 2.
\end{cases}\label{deltaGanalytics}
\end{equation}
This result applies to the disorder-averaged conductance in a wire geometry (length $L$ large compared to the width $W$), with a large number $N\gg 1$ of propagating modes, in the diffusive regime ($L$ much larger than the mean free path $l$, but much smaller than the localization length $\xi=Nl$).

An alternative way to probe the symmetry class is via the sample-to-sample fluctuations of the conductance. According to the theory of universal conductance fluctuations \cite{Alt85,Lee85}, the variance ${\rm Var}\,G$ of the conductance is proportional to $g^2/\beta$, where $g$ is the level degeneracy factor. In our case the $\beta=1\rightarrow \beta=2$ transition happens at fixed $g=2$, while the $\beta=4\rightarrow\beta=2$ transition is accompanied by $g=2\rightarrow g=1$, hence in both cases the magnetic field reduces the variance by a factor of two. The predicted values in a wire geometry are \cite{Bee97}
\begin{equation}
{\rm Var}\,G=\left(\frac{2e^2}{h}\right)^2\times\begin{cases}
2/15\rightarrow 1/15&\text{for}\;\;\beta=1\rightarrow 2,\\
1/30\rightarrow 1/60&\text{for}\;\;\beta=4\rightarrow 2.
\end{cases}\label{varGanalytics}
\end{equation}

For these quantum interference effects the crossover to $\beta=2$ happens when the magnetic flux through the wire becomes larger than a flux quantum $h/e$. Which of the two transitions applies, $\beta=1\rightarrow\beta=2$ or $\beta=4\rightarrow\beta=2$, depends on whether the supercell symmetry breaking term $t$ is small or large compared to the Thouless energy $E_{\rm T}=(h/e^2)G\delta$. In a diffusive multimode wire $G\gg e^2/h\Rightarrow E_{\rm T}\gg\delta$, hence the range of $t$ governed by the supercell symmetry is much larger for the conductance, when we need $t\ll E_{\rm T}$, than it is for the level repulsion, when the condition is $t\ll \delta$.

\section{Numerical results}
\label{sec_numerics}

\begin{figure}[tb]
\centerline{\includegraphics[width=0.9\linewidth]{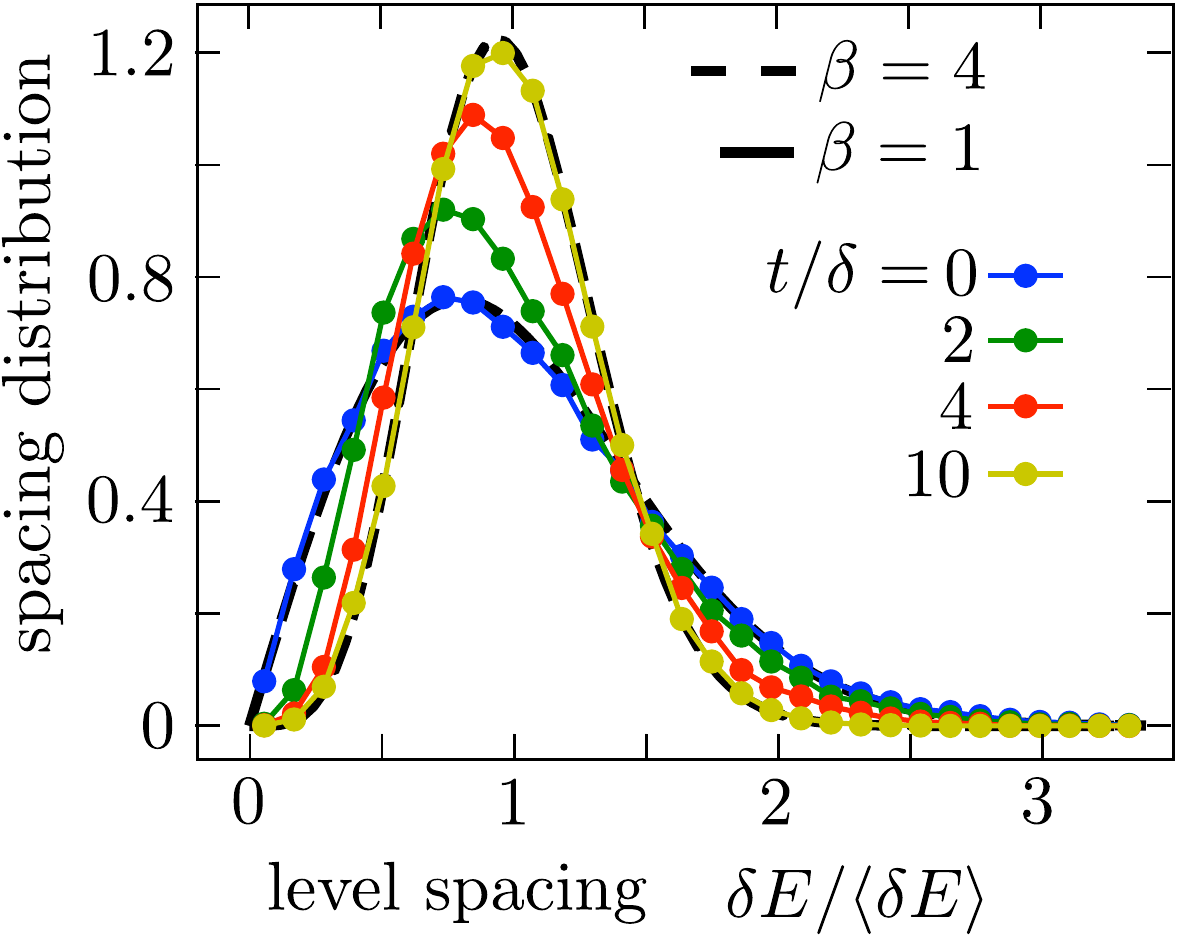}}\medskip
\centerline{\includegraphics[width=0.9\linewidth]{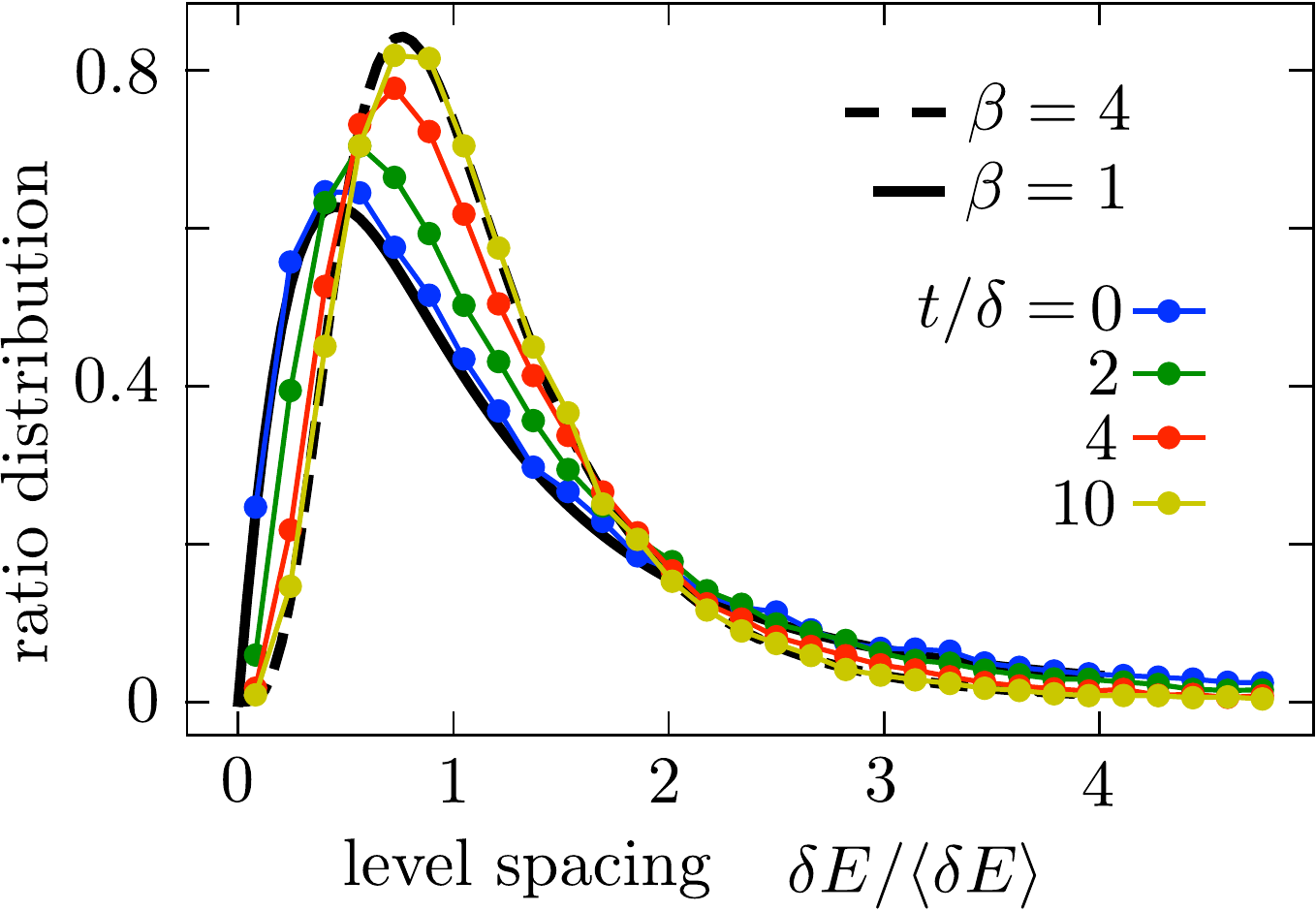}}
\caption{Top panel: Same as Fig.\ \ref{fig_GOE}, but now for a 2D square lattice (size $200\times 200$, disorder strength $V_0=0.5\,v/a_0$) and for four values of the spin-independent hopping energy $t$. The bottom panel shows the corresponding ratio distribution (with the $\beta=1$ and $\beta=4$ limits from Ref.\ \onlinecite{Ata13}).
}
\label{fig_GOE2D}
\end{figure}

\begin{figure}[tb]
\centerline{\includegraphics[width=0.9\linewidth]{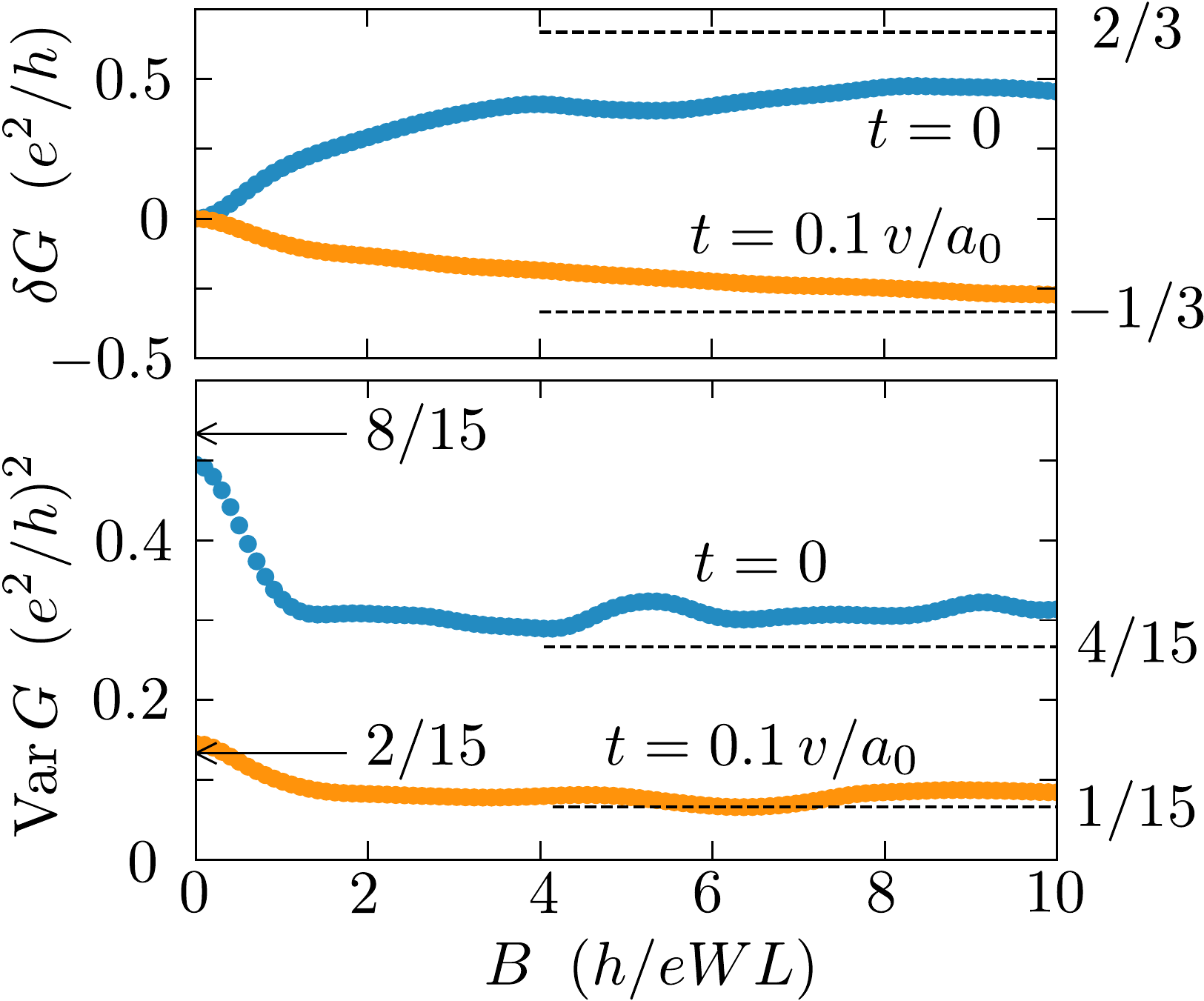}}
\caption{Magnetic field dependence of the conductance mean $\delta G=\langle G(B)\rangle-\langle G(0)\rangle$ (top panel) and conductance variance ${\rm Var}\,G=\langle G(B)^2\rangle-\langle G(B)\rangle^2$ (bottom panel), averaged over disorder in a conducting wire (length $L=1000$, width $W=200$, disorder strength $V_0=0.5\,v/a_0$, Fermi energy $E=0.2\,v/a_0$). The blue data points are in the presence of the supercell symmetry ($t=0$), for the gold data points the symmetry is broken ($t=0.1\,v/a_0$). The arrows and dashed lines indicate the analytical predictions \eqref{deltaGanalytics} and \eqref{varGanalytics} in the limit $N\rightarrow\infty$.
}
\label{fig_conductance}
\end{figure}

We have studied the effect of the supercell symmetry numerically, using the \textit{Kwant} tight-binding package \cite{kwant,zenodo}. For computational efficiency we took a 2D square lattice, rather than a 3D lattice, given by the Hamiltonian
\begin{align}
H={}& v(\sigma_x\sin a_0p_x+\sigma_y\sin a_0p_y)\nonumber\\
&+t\sigma_0 (\cos a_0p_x+\cos a_0p_y)+V(\bm{r})\sigma_0. \label{Hdef2D}
\end{align}
The random potential $V$ was chosen independently on each site, uniformly in the interval $(-V_0/2,V_0/2)$. 

For the level statistics we took a square geometry,\footnote{In all our systems we truncate the lattice without applying periodic boundary conditions. The parity of the number of lattice sites then does not matter.}
 on a lattice of size $200\,a_0\times 200\,a_0$. We calculated the distribution of the nearest-neigbor spacings of the twofold degenerate levels in the interval $|E-0.2\,v/a_0|<4\cdot 10^{-3}\,v/a_0$ (mean level spacing $\delta=3.56\cdot 10^{-4}\,v/a_0$, approximately constant in this energy range), averaging over some 2000 disorder realizations. Note that the disorder potential breaks chiral symmetry,\footnote{Chiral symmetry means that the Hamiltonian $\sigma_x\sin p_x+\sigma_y\sin p_y$ anticommutes with $\sigma_z$, enforcing a $\pm E$ symmetry in the spectrum. This symmetry plays no role in our analysis, because it is broken by the $V\sigma_0$ disorder potential.} so there is no $\pm E$ symmetry in the spectrum.
 
As an extra check, we also calculated the ratio distribution \cite{Ata13}, meaning the probability distribution $P(r)$ of the ratio $r_n=s_n/s_{n-1}$ of two consecutive level spacings $s_n=E_{n+1}-E_n$.

For the conductance we took a disordered wire of width $W=200\,a_0$ and length $L=1000\,a_0$. The end points are connected to heavily doped metal leads, modelled on the lattice by breaking the transverse bonds. The transmission matrix $\bm{t}$ at Fermi energy $E$ determines the zero-temperature two-terminal conductance $G=(e^2/h)\,{\rm Tr}\,\bm{tt}^\dagger$. We took $E=0.2\,v/a_0$, when the number of propagating modes through the disordered region equals $N=52$ (counting degeneracies). The mean free path for $V_0=0.5\,v/a_0$ is estimated at $l=150\,a_0$, from the Drude formula $G\approx (Ne^2/h)(1+L/l)^{-1}$. The localization length $\xi=Nl$ is then larger than $L$, so we are in the diffusive regime.

Fig.\ \ref{fig_GOE2D} shows the transition from the $\beta=1$ to $\beta=4$ level spacing and ratio distributions. The transition from weak localization to weak anti-localization is shown in Fig.\ \ref{fig_conductance}, as well as the transition from $\beta=1$ to $\beta=4$ conductance fluctuations. It is difficult to fully reach the large-$N$ regime where the analytical results \eqref{deltaGanalytics} and \eqref{varGanalytics} apply, so the agreement analytics--numerics remains qualitative for the conductance.

\begin{figure}[tb]
\centerline{\includegraphics[width=0.9\linewidth]{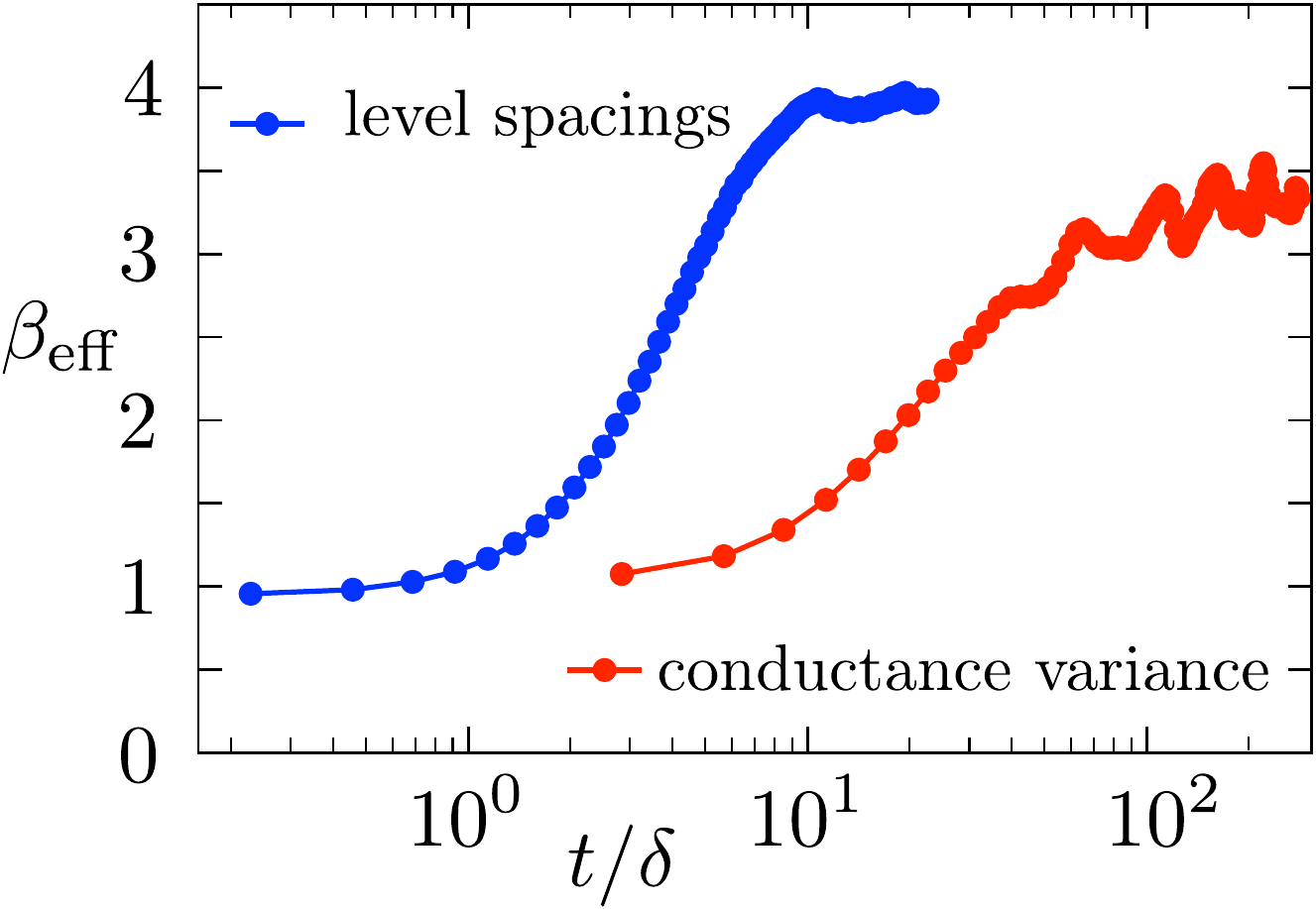}}
\caption{Transition from $\beta=1$ to $\beta=4$ with increasing spin-independent hopping energy $t$, as measured via the level spacing distribution (red data points, same parameters as in Fig.\ \ref{fig_GOE2D}) or via the variance of the conductance (blue data points, same parameters as in Fig.\ \ref{fig_conductance}, at $B=0$). The transition is quantified by an effective parameter $\beta_{\rm eff}$. For the conductance this is defined by $\beta_{\rm eff}=\tfrac{8}{15}(e^2/h)^2({\rm Var}\,G)^{-1}$. For the level spacing we fitted the data to the Wigner surmise interpolation \cite{Mod14} $P(s)=c s^{\beta_{\rm eff}} \exp(-c' s^2)$, with $s=\delta E/\delta$ and coefficients $c,c'$ such that the zeroth and first moments of $P(s)$ are equal to unity.
}
\label{fig_scaling}
\end{figure}

In Fig.\ \ref{fig_scaling} we show that the effect of the supercell symmetry is suppressed more rapidly by the spin-independent hopping energy $t$ if we consider the level spacings (when we need $t\gtrsim\delta$) than it is if we consider the conductance (when we need $t\gtrsim E_{\rm T}$). In the conductance calculations $G\approx 7e^2/h\Rightarrow E_{\rm T}/\delta\approx 7$, so we expect about an order of magnitude difference in the onset of the two transitions, in accord with Fig.\ \ref{fig_scaling}.

\section{Conclusion}
\label{conclude}

In summary, we have identified a supercell symmetry and a resulting ``fake'' time-reversal symmetry operation, squaring to $+1$ rather than $-1$, which explains the $\beta=1$ spectral statistics of the Kramers-Weyl Hamiltonian \eqref{Hdef} in the absence of the spin-independent hopping term $\propto t\cos p$. The same symmetry is responsible for the appearance of weak localization in the magnetoconductance.

The crossover from $\beta=1$ to $\beta=4$ level repulsion happens quickly, when $t$ becomes larger than the mean level spacing $\delta$. The crossover from weak localization to weak antilocalization happens at larger $t$, larger by a factor of conductance $G\times h/e^2$. This delayed crossover in the magnetoconductance may make the effect of the supercell symmetry more easily observable.

A similar shift of symmetries has been observed when comparing two discretization schemes of lattice Dirac operators on a torus \cite{Kie14,Kie17}. The Dirac Hamiltonian $-i\bm{\nabla}\cdot\bm{\sigma}$ needs a special ``staggered'' discretization of the spatial derivative to make sure that the low-energy states are only near $\bm{p}=0$. The ``naive'' discretization $\partial f/\partial x \mapsto (2a)^{-1}[f(x+a)-f(x-a)]$ introduces an additional Dirac cone at $\bm{p}=\pi/a$ (fermion doubling \cite{Nie81,Kap09}). 

If one then imposes periodic boundary conditions, the naive discretization obeys the supercell symmetry \eqref{supercellsymm} if the number of lattice sites is even but not if it is odd. The way this works out for the spectral statistics is different in Refs.\ \cite{Kie14,Kie17} than it is here, because of the presence of chiral symmetry, but the mechanism is the same.

\acknowledgments

We have benefited from discussions with Anton Akhmerov. This project has received funding from the Netherlands Organization for Scientific Research (NWO/OCW) and from the European Research Council (ERC) under the European Union's Horizon 2020 research and innovation programme.

\end{document}